\documentclass[12pt,cite,epsf,epsfig]{article}
\usepackage{epsfig}

\begin{document}

\author{S. Dev\thanks{dev5703@yahoo.com}, Shivani Gupta\thanks{shiroberts\_1980@yahoo.co.in} and Radha Raman Gautam\thanks{gautamrrg@gmail.com}}

\title{Broken $S_3$ Symmetry in the Neutrino Mass Matrix}
\date{\textit{Department of Physics, Himachal Pradesh University, Shimla 171005, India.}\\
\smallskip}

\maketitle
\begin{abstract}
We explore the feasibility of the discrete flavor symmetry $S_3$ to explain the pattern of neutrino masses and mixings. In the flavor basis, the neutrino mass matrix is taken to be invariant under $S_3$ symmetry at the zeroth order. The effects of breaking $S_3$ symmetry by matrices invariant under different $S_2$ subgroups of $S_3$ are studied. The resulting retrocirculant perturbation matrix leads to the perturbed $S_3$ neutrino mass matrix having a trimaximal eigenvector as suggested by the solar neutrino data. It is found that after the third order perturbation the neutrino mixings only depend on the perturbation parameter and are consistent with the current experimental data leading to very small deviations from tribimaximal mixing. These perturbations in the $S_3$ invariant neutrino mass matrix result in interesting interplay between the solar and the reactor neutrino mixing angles. We also get a strongly suppressed range of effective Majorana mass which lies well within the reach of the ongoing experiments. 
\end{abstract}

\section{Introduction}
The recent neutrino oscillation data have greatly enriched our knowledge of the flavor structure of leptons. After the discovery of neutrino oscillations, there has been a considerable progress in determining the values of the neutrino mass-squared differences and mixing angles relating the mass eigenstates to the flavor eigenstates. The remaining challenges are to determine the absolute mass scale of neutrinos, the mass hierarchy in the neutrino sector (sign of $\triangle m_{23}^2$) and the value of 1-3 mixing angle (in particular if it is zero or not) amongst others. One has to understand the systematics of the experimental data to unravel symmetries of the lepton mass matrices. A successful phenomenological Ans\"{a}tz for the neutrino mixing matrix consistent with the present neutrino data was proposed by Harrison, Perkins and Scott \cite{1} known as tribimaximal mixing (TBM) and is given by
\begin{equation}
U_{TBM} = \left(
\begin{array}{ccc}
\frac{-2}{\sqrt{6}} & \frac{1}{\sqrt{3}} & 0 \\
\frac{1}{\sqrt{6}}&
\frac{1}{\sqrt{3}}& \frac{-1}{\sqrt{2}}\\
\frac{1}{\sqrt{6}}&
\frac{1}{\sqrt{3}}& \frac{1}{\sqrt{2}}\\
\end{array}
\right).
\end{equation} 
Recently, non-Abelian flavor symmetries \cite{2, 3, 5} have been invoked intensively to account for TBM. These non-Abelian flavor symmetries can explain various, apparently unrelated, phenomena in flavor physics thus providing hints for the new physics beyond the Standard Model (SM). The basic features of the neutrino masses and mixings can be traced back to a minimum realization of the smallest discrete non-Abelian group $S_3$ which is the permutation group of three objects. A large number of papers \cite{3, 5} have presented detailed models based on $S_3$ symmetry. The permutation matrices in the three dimensional reducible representation are
\begin{equation}
S^{(1)} = \left(
\begin{array}{ccc}
1 &0& 0 \\
0&1&0\\
0&0&1\\
\end{array}
\right),
\end{equation} 
\begin{equation}
S^{(123)} = \left(
\begin{array}{ccc}
0 &0& 1 \\
1&0&0\\
0&1&0\\
\end{array}
\right),
S^{(132)}=\left(
\begin{array}{ccc}
0&1&0\\
0&0&1\\
1&0&0\\
\end{array}
\right),
\end{equation} 
\begin{equation}
S^{(12)}=\left(
\begin{array}{ccc}
0 &1& 0 \\
1&0&0\\
0&0&1\\
\end{array}
\right),
S^{(13)}=\left(
\begin{array}{ccc}
0 &0& 1 \\
0&1&0\\
1&0&0\\
\end{array}
\right),
S^{(23)} = \left(
\begin{array}{ccc}
1&0&0\\
0&0&1\\
0&1&0\\
\end{array}
\right),
\end{equation}
where the matrices in each equation belong to the same class of $S_3$. 
The TBM mixing matrix $U_{TBM}$ is one of the transformations which relates the given basis to the irreducible one thus implying a close connection between TBM and $S_3$ symmetry. The neutrino mass matrix $M_\nu$ is invariant under $S_3$ which requires
\begin{equation}
[S,M_\nu]=0,
\end{equation}
where $S$ is any of the six permutation matrices given in Eqns. (2, 3, 4). In the present work, we consider the possibility in which the neutrino mass matrix is dominated by a term with $S_3$ symmetry. We then consider possible perturbations that violate this symmetry so that the perturbed neutrino mass matrix $(M_\nu')$ still satisfies the $S_3$ invariant constraint \cite{7}
\begin{equation}
M_{\nu_{(\beta \beta)}}' - M_{ \nu_{(\gamma \gamma)}}' = M_{ \nu_{(\delta \gamma )}}' - M_{\nu_{(\delta \beta)}}', (\beta \neq \gamma \neq \delta).
\end{equation}
Such connections have been considered in the literature from phenomenological point of view. In this analysis, we take the charged lepton mass matrix to be diagonal. This simple choice of basis can be made as the physical mixing matrix does not depend on the choice of basis. The property of basis independence allows one to rotate the charged lepton mass matrix to the flavor basis where it is diagonal. If a horizontal symmetry exists it must be simultaneously a symmetry of the charged leptons as well as the neutrinos before the gauge symmetry breaking. After symmetry breaking when the fermions acquire mass, the charged lepton and the neutrino mass matrices should be constrained by different representations of the symmetry group in order to have non-zero mixing. The full group before symmetry breaking is at least a product group of the symmetry groups of the charged leptons and the neutrinos. Here we consider $S_3$ to be the residual group in the neutrino sector. For the charged lepton sector, $Z_3$ symmetry can be taken as the residual symmetry which yields non degenerate diagonal charged lepton mass matrix \cite{4}.
\section{$S_3$ invariant neutrino mass matrix}
The neutrino mass matrix implied by the commutator in Eqn. (5) is given by 
\begin{equation}
M_\nu= aI+bD.\\
\end{equation}
where
\begin{equation}
I=  \left(
\begin{array}{ccc}
1 &0& 0 \\
0&1&0\\
0&0&1\\
\end{array}
\right), 
D= \left(
\begin{array}{ccc}
1&1&1\\
1&1&1\\
1&1&1\\
\end{array}
\right)
\end{equation}

where $a$ and $b$ are in general complex and $D$ is called the democratic matrix. The neutrino mass matrix can be diagonalized by the TBM matrix $U_{TBM}$. The eigenvalues of $M_{\nu}$ are $a$, $a+3b$ and $a$ respectively. Here, the two degenerate eigenvalues correspond to the mass eigenvalues $m_1$ and $m_3$ contrary to the experimental data where $m_1$ and $m_2$ have smaller mass difference. This problem was overcome by Jora \textit{et al.} \cite{5} by introducing a Majorana-type $CP$ violating phase $\alpha$ between the complex vectors $a$ and $a+3b$ which does not affect the usual neutrino oscillations but affects the rate of neutrinoless double beta decay. The phase $\alpha$ can be adjusted to ensure equal magnitude (but different directions) of $a$ and $a+3b$, thereby, leading to degenerate mass spectrum for $S_3$ invariant $M_\nu$ at the zeroth order. In the present work, the complex plane is oriented in a manner so that the parameter $b$ is completely imaginary. The complex vector $a$ lies in the fourth quadrant and is given by
\begin{equation}
a=|a|e^{-i \alpha /2}.
\end{equation}
\begin{figure}
\begin{center}
\epsfig{file=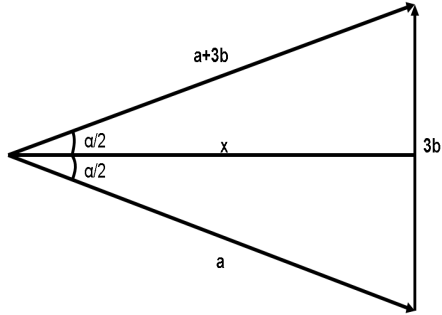,height=6.0cm,width=6.0cm}
\end{center}
\caption{Angle $\alpha$ between two equal length vectors $a$ and $a+3b$.}
\end{figure} 
and the allowed range of $\alpha$ is $0 \leq \alpha < \pi$. The complex parameters $a$ and $b$ can be written in terms of the real free parameter $x$ as
\begin{eqnarray}
|a|=x \sec (\alpha/2),\nonumber \\
|b|=\frac{2}{3}x \tan (\alpha/2).
\end{eqnarray}
For the Majorana-type $CP$ violating phase $\alpha=180^o$, the parameter $x$ becomes zero and this value of $\alpha$ is, hence, not allowed.
Democratic charged lepton mass matrices of the form $bD$ and diagonal neutrino mass matrices of the form $aI$ have been discussed earlier \cite{6} in the context of $S_3$ symmetry. The possibility $a=0$ has, also been discussed in the literature \cite{7}. $S_3$ flavor symmetry in the context of Type (I+II) see saw has been discussed in ref. \cite{8}. A generic feature of this approach is a democratic charged lepton mass matrix and a diagonal neutrino mass matrix at the zeroth order. In contrast, we consider the most general $S_3$ invariant neutrino mass matrices at the zeroth order in the flavor basis where the charged lepton mass matrix is diagonal.
\section{Perturbations from $S_3$}
In this section, we study the effects of perturbations from $S_3$ invariant neutrino mass matrix $M_\nu$, by the elements of $S_3$ group itself. We follow a simple and highly predictive approach to obtain the results of perturbations order by order such that the neutrino mass matrix $M_\nu$ is invariant under $S_3$ at the zeroth order. Parametrizing higher order perturbations of decreasing importance in terms of a small real dimensionless perturbation parameter $\lambda$, the perturbed neutrino mass matrix $M'_\nu $ is given by
\begin{equation}
M'_\nu = M_\nu +\mu (\lambda [S^p]+\lambda^2 [S^q]+\lambda^3 [S^r]), 
\end{equation}
where $M_\nu$ is invariant under $S_3$ and  $S^p$, $S^q$ and $S^r$ can be any of $S^{(12)}$, $S^{(13)}$ and $S^{(23)}$ such that $p \neq q \neq r$ and $\mu$ is the real parameter with dimensions of mass and magnitude of the order of one in the units of the absolute mass scale. The perturbation matrices are given in Eqn. (4) and belong to the same class of $S_3$. Each perturbation matrix given above is invariant under a different $S_2$ subgroup of $S_3$. The advantage of this approach of perturbation is that there are only four free parameters. Another appealing aspect of this approach is that the $S_3$ symmetry is broken by its elements in the same representation. We see that even after breaking $S_3$ by above matrices, the $S_3$ invariant constraint (Eqn. 6) remains intact which leads to a trimaximal eigenvector for the perturbed neutrino mass matrix $M'_\nu$.  We follow the line of argument given in Ref. [5] to decide as to which of the three perturbation matrices $S^{(12)}$, $S^{(13)}$ and $S^{(23)}$ is the favoured one to be the first order perturbation. Since zeroth order neutrino mass matrix $M_\nu$ has two degenerate eigenvalues, we use quantum mechanical degenerate perturbation theory. Due to the double degeneracy between the eigenvectors corresponding to $m_1$ and $m_3$, the tribimaximal mixing matrix $U_{TBM}$ given in Eqn. (1) is not the unique one to diagonalize $M_\nu$. Rather it is a more general matrix $U_{TBM} R(\phi _{13})$ where $R(\phi _{13})$ is the rotation matrix in the 1-3 plane given by
\begin{equation}
R(\phi _{13}) = \left(
\begin{array}{ccc}
\cos \phi _{13} &0& -\sin \phi _{13} \\
0&1&0\\
\sin \phi _{13}&0&\cos \phi _{13}\\
\end{array}
\right).
\end{equation}
In the ordinary quantum mechanical perturbation theory, the first order correction to the $i^{th}$ eigenvector is given by $\frac{\langle i | H^{(1)} | j \rangle }{E_i -E_j}$. However, in case of degenerate perturbation theory, when $E_i = E_j$, the numerator has to vanish which implies the following condition for the (1,3) matrix element:
\begin{equation}
(R(\phi _{13})^T U_{TBM}^T S^{p} U_{TBM}R(\phi _{13}))_{13}=0.
\end{equation}
Amongst the three perturbation matrices only $S^{(23)}$ satisfies the above condition thus leading to $U_{TBM}$ for $\phi=0$. The solution $\phi=\pi$, on the other hand, just changes the sign of first and third column of tribimaximal mixing matrix $U_{TBM}$. However, the other two solutions $\phi=\frac{\pi}{2}$ and $\phi=\frac{3\pi}{2}$ do not lead to the desired TBM form. The other two matrices $S^{(12)}$ and $S^{(13)}$ also do not lead to the desired form of TBM mixing matrix, thus, making $S^{(23)}$ as the favoured perturbation at the first order. The neutrino mass matrix after the first order perturbation with $S^{(23)}$ becomes 
\begin{equation}
M^{(1)}_\nu = \left(
\begin{array}{ccc}
a+b+ \mu \lambda & b & b \\
b & a+b & b+ \mu  \lambda \\
b & b+ \mu \lambda & a+b \\
\end{array}
\right).
\end{equation}
The eigenvalues of this mass matrix are
\begin{center} 
\begin{eqnarray}
\begin{array}{ccc}
m_1=a+ \mu \lambda ,\\
m_2=a+3b+ \mu \lambda ,\\
m_3=a- \mu \lambda .\\
\end{array}
\end{eqnarray}
\end{center} 
Here, we see that $|m_1|=|m_2|$ and $|m_3|<|m_1|$ when $\lambda$ is positive i.e. after the first order perturbation $|m_1|$ and $|m_2|$ remain degenerate, while $|m_3|<|m_1|$ thus implying that our perturbation is in the right direction as experimentally the mass-squared difference $\Delta m_{12}^2$  is much smaller than $\Delta m_{13}^2$. The neutrino mixing matrix after this first order perturbation is still $U_{TBM}$. The next step is to add the higher order perturbations to remove the degeneracy in the magnitudes of $m_1$, $m_2$ and also to deviate from TBM. This can be done by taking $\lambda ^2 S^{(12)}$ as the second order and $\lambda ^3 S^{(13)}$ as the third order perturbation or $\lambda ^2 S^{(13)}$ as the second order and $\lambda ^3 S^{(12)}$ as the third order perturbation. We investigated both these possibilities in our analysis and found that the choice of the second and the third order perturbation matrices only affects the predictions for the 2-3 mixing angle which we discuss later. Taking $\lambda ^2 S^{(12)}$ as the second order perturbation and $\lambda ^3 S^{(13)}$ as the third order perturbation, the perturbation matrix becomes
\begin{equation}
\mu \left(
\begin{array}{ccc}
\lambda & \lambda^2 & \lambda^3 \\
\lambda^2& \lambda^3 & \lambda \\
\lambda^3 & \lambda & \lambda^2 \\
\end{array}
\right).
\end{equation}
Hence, the perturbed neutrino mass matrix becomes
\begin{equation}
M'_\nu = \left(
\begin{array}{ccc}
a+b+ \mu \lambda & b+\mu \lambda^2 & b+\mu \lambda^3 \\
b+\mu \lambda^2 & a+b+\mu \lambda^3& b+\mu \lambda \\
b+\mu \lambda^3 & b+\mu \lambda & a+b+\mu \lambda^2 \\
\end{array}
\right).
\end{equation}
It is interesting to note that the perturbation matrix in Eqn. (16) is retrocirculant having a trimaximal eigenvector $(\frac{1}{\sqrt{3}},\frac{1}{\sqrt{3}},\frac{1}{\sqrt{3}})^T$ \cite{9, 10}. Also the perturbed neutrino mass matrix $M'_\nu$ is a magic mass matrix \cite{10, 11}. The eigenvalues of the perturbed neutrino mass matrix $M'_\nu$ upto the third order perturbation are given by
\begin{equation}
\begin{array}{ccc}
m_1 \approx x \sec (\frac{\alpha}{2}) e^{-i \alpha /2} +\mu( \lambda - \frac{\lambda ^2}{2} - \frac{\lambda ^3}{8}),\\
m_2 = x \sec (\frac{\alpha}{2}) e^{-i \alpha /2} + 2ix \tan (\frac{\alpha}{2}) + \mu (\lambda + \lambda^2 + \lambda^3),\\
m_3 \approx x \sec (\frac{\alpha}{2}) e^{-i \alpha /2}+ \mu (- \lambda + \frac{\lambda ^2}{2} + \frac{\lambda ^3}{8}).
\end{array}
\end{equation}
The magnitudes of the eigenvalues which are the neutrino masses are given by
\begin{equation}
\begin{array}{ccc}
|m_1| \approx  x \sec \frac{\alpha}{2}[ 1+\cos ^2 \frac{\alpha}{2} \lbrace \frac{\lambda \mu}{x}+\frac{\lambda ^2 (\mu^2-x \mu- \mu^2 \cos ^2 \frac{\alpha}{2})}{2x^2}- \frac{\lambda ^3 (x^2 \mu + 4x \mu^2 +4 \mu^3 \cos ^2 \frac{\alpha}{2}-4x \mu^2 \cos ^2 \frac{\alpha}{2}-4 \mu^3 \cos ^4 \frac{\alpha}{2})}{8x^3}\rbrace ],\\

|m_2| \approx  x \sec \frac{\alpha}{2}[ 1+\cos ^2 \frac{\alpha}{2} \lbrace \frac{\lambda \mu}{x}+\frac{\lambda ^2 (\mu^2+ 2x \mu-\mu^2 \cos ^2 \frac{\alpha}{2})}{2x^2}+ \frac{\lambda ^3 (2x^2 \mu +2x \mu^2 - \mu^3 \cos ^2 \frac{\alpha}{2}-2x \mu^2 \cos ^2 \frac{\alpha}{2}+ \mu^3 \cos ^4 \frac{\alpha}{2})}{2x^3}\rbrace ],\\

|m_3| \approx  x \sec \frac{\alpha}{2}[ 1+\cos ^2 \frac{\alpha}{2} \lbrace \frac{-\lambda \mu}{x}+\frac{\lambda ^2 (\mu^2 +x \mu -\mu^2 \cos ^2 \frac{\alpha}{2})}{2x^2}+ \frac{\lambda ^3 (x^2 \mu -4x \mu^2 +4 \mu^3 \cos ^2 \frac{\alpha}{2}+4x \mu^2 \cos ^2 \frac{\alpha}{2}-4 \mu^3 \cos ^4 \frac{\alpha}{2})}{8x^3}\rbrace ].
\end{array}
\end{equation}
These neutrino masses are made real positive by the phase matrix
\begin{equation}
P=\left(
\begin{array}{ccc}
e^{-i \tau} & 0 & 0 \\
0 & e^{-i \sigma} & 0 \\
0 & 0 & e^{-i \rho} \\
\end{array}
\right)
\end{equation}
where
\begin{equation}
\begin{large}
\begin{array}{ccc}
\tau \approx \frac{1}{2}\tan^{-1}(\frac{- \tan (\frac{\alpha}{2})}{1+ \frac{\mu}{x}(\lambda - \frac{\lambda ^2}{2 }- \frac{\lambda ^3}{8})}),\\
\sigma \approx \frac{1}{2}\tan^{-1}(\frac{ \tan (\frac{\alpha}{2})}{1+ \frac{\mu}{x}(\lambda + \lambda^2 + \lambda^3)}),\\
\rho \approx \frac{1}{2}\tan^{-1}(\frac{- \tan (\frac{\alpha}{2})}{1+ \frac{\mu}{x}(- \lambda + \frac{\lambda ^2}{2}+ \frac{\lambda ^3}{8})}).
\end{array}
\end{large}
\end{equation}

It can be seen from Eqn. (19) that the neutrino masses depend on the Majorana-type $CP$ violating phase $\alpha$. However, the lepton number conserving neutrino oscillations do not depend on this phase. Therefore, we find the mass-squared differences are independent of the Majorana phase.
\begin{equation}
\begin{array}{cc}
\triangle m_{12}^2 \approx 3x\mu\lambda ^2 + (3\mu^2+\frac{9x \mu}{4 })\lambda ^3,\\
\triangle m_{23}^2 \approx 4x \mu \lambda + x \mu \lambda ^2+(3 \mu^2+\frac{7x \mu}{4 })\lambda ^3.
\end{array}
\end{equation}
 Notice that the leading order term in the solar mass-squared difference $\triangle m_{12}^2$ is of the order of $\lambda ^2$ which makes it naturally small to be consistent with the experimental data. The neutrino mixing matrix after the third order perturbation is given by
\begin{equation}
U' = \left(
\begin{array}{ccc}
U'_{11} & U'_{12} & U'_{13} \\
U'_{21} & U'_{22} & U'_{23} \\
U'_{31} & U'_{32} & U'_{33} \\
\end{array}
\right)
\end{equation}
where
\begin{eqnarray}
U'_{11} \approx  -\sqrt{\frac{2}{3}}+\frac{1}{16}\sqrt{\frac{3}{2}} \lambda^2  - \frac{1}{16}\sqrt{\frac{3}{2}}  \lambda^3 , \nonumber \\
U'_{12} \approx  \frac{1}{\sqrt{3}}, \nonumber\\
U'_{13} \approx  \frac{\lambda}{2\sqrt{2} }-\frac{\lambda ^2}{4\sqrt{2} }-\frac{\lambda ^3}{64\sqrt{2}}, \nonumber\\
U'_{21} \approx  \sqrt{\frac{1}{6}}-\frac{1}{4}\sqrt{\frac{3}{2}}\lambda + \frac{3}{32}\sqrt{\frac{3}{2}} \lambda^2 +\frac{5}{128}\sqrt{\frac{3}{2}} \lambda^3,\nonumber \\
U'_{22} \approx  \frac{1}{\sqrt{3}}, \nonumber\\
U'_{23} \approx -\frac{1}{\sqrt{2}}-\frac{\lambda}{4\sqrt{2} }+\frac{7\lambda ^2}{32\sqrt{2}}-\frac{11\lambda ^3}{128\sqrt{2}}, \nonumber\\
U'_{31} \approx  \sqrt{\frac{1}{6}}+\frac{1}{4}\sqrt{\frac{3}{2}}\lambda - \frac{5}{32}\sqrt{\frac{3}{2}} \lambda^2 +\frac{3}{128}\sqrt{\frac{3}{2}} \lambda^3, \nonumber\\
U'_{32} \approx  \frac{1}{\sqrt{3}}, \nonumber\\
U'_{33} \approx  \frac{1}{\sqrt{2}}-\frac{\lambda}{4\sqrt{2} }+\frac{\lambda ^2}{32\sqrt{2} }+\frac{13\lambda ^3}{128\sqrt{2}}.
\end{eqnarray}
An interesting feature of this type of breaking of $S_3$ is that the neutrino mixing matrix is completely determined from the retrocirculant perturbation matrix and is, thus, independent of the parameters appearing in the zeroth order neutrino mass matrix $M_\nu$. As a consequence, the deviations from TBM are entirely determined in terms of single unknown perturbation parameter $\lambda$. The neutrino mixing angles upto the third order perturbation are given by
\begin{equation}
\sin\theta_{12}^2 \approx \frac{1}{3}+\frac{\lambda ^2}{24 }-\frac{\lambda ^3}{24 },
\end{equation}
\begin{equation}
\sin\theta_{13}^2 \approx \frac{\lambda ^2}{8 }-\frac{\lambda ^3}{8 },
\end{equation}
and
\begin{equation}
\sin\theta_{23}^2 \approx \frac{1}{2}+\frac{\lambda}{4 }-\frac{\lambda ^2}{8 }.
\end{equation}
The solar and reactor neutrino mixing angles are related by 
\begin{equation}
\sin\theta_{13}^2=3\sin\theta_{12}^2-1.
\end{equation}
As pointed out earlier in this section, the second and third order perturbation matrices can be interchanged. As a consequence of this interchange, the predictions for all the parameters remain intact except for the 2-3 mixing angle $\theta _{23}$ which is given by 
\begin{equation}
\sin\theta_{23}^2 \approx \frac{1}{2}-\frac{\lambda}{4 }+\frac{\lambda ^2}{8 }.
\end{equation}
It can be seen that the 2-3 mixing angle in this case is shifted to below maximal by the same amount as it was above maximal in the earlier case.

\section{Numerical results}
The above results are obtained when the perturbation parameter $\lambda$ is positive. Since $\lambda $ is real it can also be negative. When $\lambda$ is positive, we get the inverted hierarchical mass spectrum for neutrinos and a negative $\lambda$ gives a normal hierarchical mass spectrum. We use the constraints for the mass squared differences from the known oscillation data \cite{12} viz.
\begin{eqnarray}
\Delta m_{12}^{2} &=&7.67_{(-0.19,-0.36,-0.53)}^{(+0.16,+0.34,+0.52)}\times
10^{-5}eV^{2}, \nonumber \\
\Delta m_{23}^{2} &=&\pm
2.39_{(-0.8,-0.20,-0.33)}^{(+0.11,+0.27,+0.47)}\times 10^{-3}eV^{2}. \nonumber \\ 
\end{eqnarray}
For normal hierarchy \textit{i.e.} for negative $\lambda$, the parameters $\mu$ and $\lambda$ are constrained to lie in the range $(0.2-0.3)eV$ and $((-0.037)-(-0.095))$  (Fig. 2(b)) respectively at the $3\sigma$ C.L. in this model whereas the free parameter $x$ and the Majorana phase $\alpha$ are constrained to the range $(0.035-0.09)eV$ and $(0^o-160^o)$ respectively at the same confidence level. The model, also, leads to small deviations of the mixing angles from TBM, given by 
\begin{eqnarray}
\theta _{12}= 35.26^o-35.29^o, \\
 \theta _{13}= 0.8^o-2.2^o, \\
 \theta _{23}= 43.45^o-44.45^o.
\end{eqnarray}
while the allowed ranges for these mixing angles \cite{12} are given by
\begin{eqnarray}
\theta_{12}=33.96^{o(+1.16,+2.43,+3.80)}_{(-1.12,-2.13,-3.10)},\nonumber\\
\theta_{23}=43.05^{o(+4.18,+7.83,+10.32)}_{(-3.35,-5.82,-7.93)},\nonumber\\
\theta_{13}<12.38^o (3\sigma).
\end{eqnarray}\\
For inverted hierarchy \textit{i.e.} positive $\lambda$ the allowed ranges for the parameters  $\mu$ and $\lambda$ are $(1.44-0.2)eV$ and $(0.017-0.043)$  (Fig. 3(b)) respectively at the $3\sigma$ C.L. whereas the free parameter $x$ and the Majorana phase $\alpha$ are constrained to the range $(0.2-1.44)eV$ and $(0^o-160^o)$ respectively at the same confidence level. The deviations of the mixing angles are given by
\begin{eqnarray}
\theta _{12}= 35.265^o-35.269^o, \\
 \theta _{13}= 0.35^o-0.90^o, \\
 \theta _{23}= 45.25^o-45.61^o.
\end{eqnarray}
\begin{figure}
\begin{center}
\epsfig{file=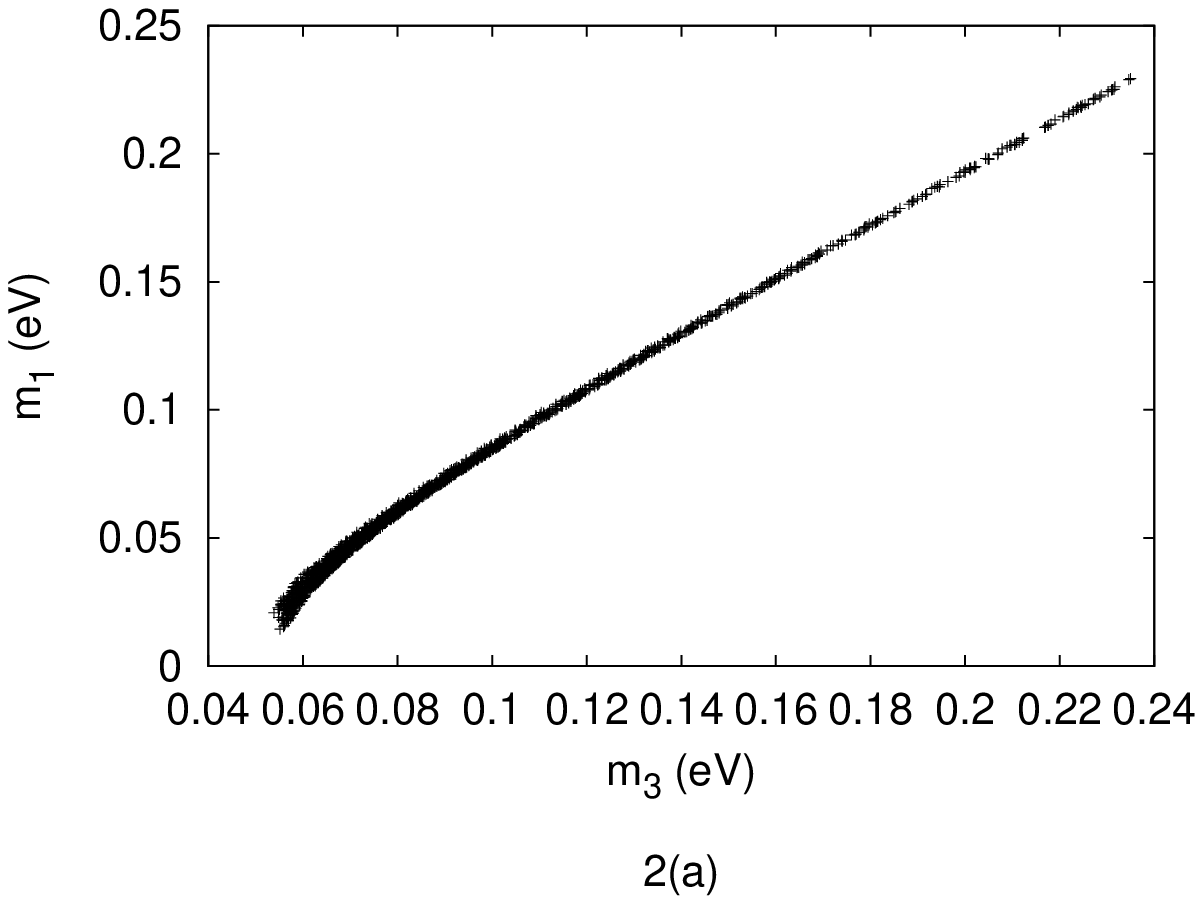,height=5.0cm,width=5.0cm}
\epsfig{file=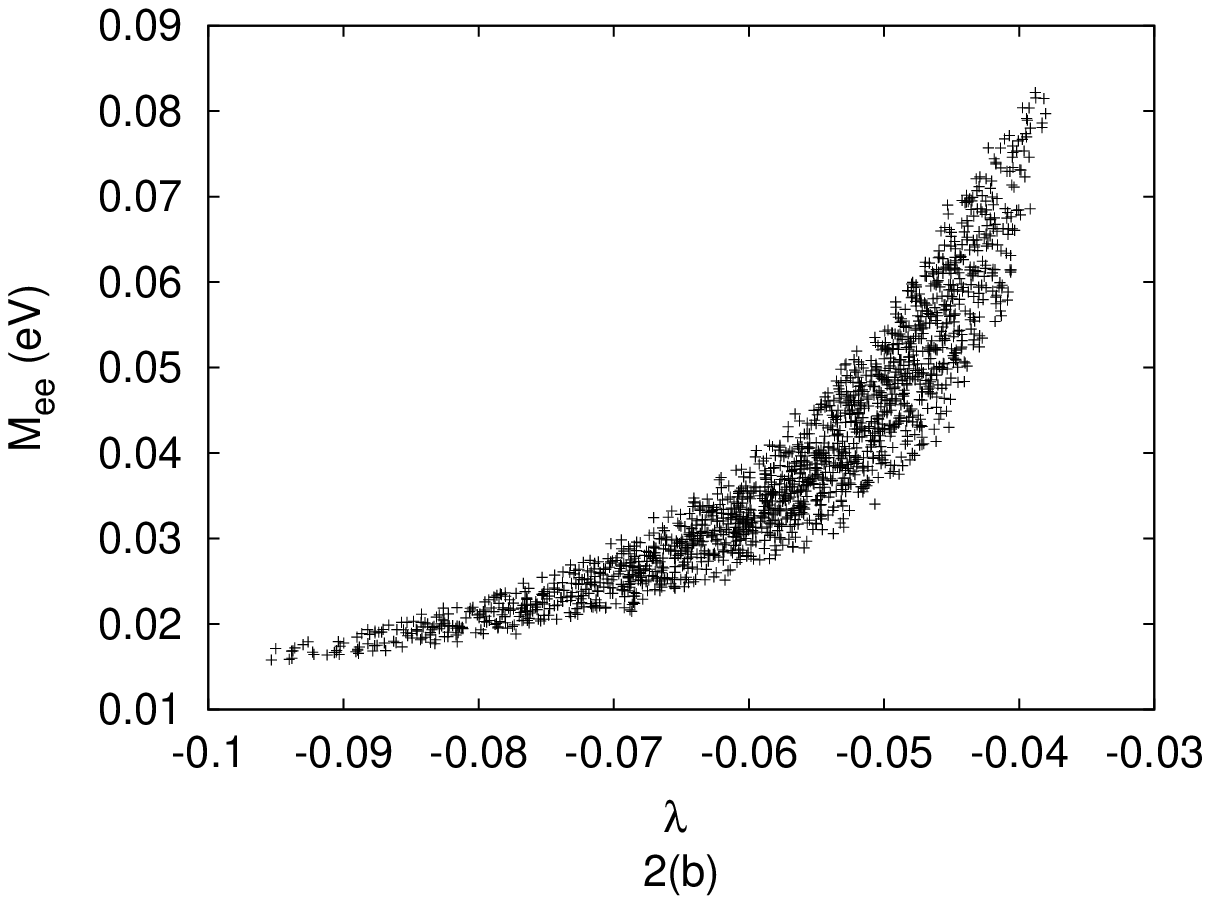,height=5.0cm,width=5.0cm}
\end{center}
\caption{Correlation plots for the normal hierarchy.}
\end{figure}  
It can be seen from the neutrino mixing matrix given in Eqn. (23) that there is no Dirac-type $CP$ violation in this model as $\lambda$ is taken to be real. The $CP$ violation in this model is only due to the phase $\alpha$ which gives Majorana-type $CP$ violation. However, $\lambda$ may, in general, be complex and there will be another $CP$ violating phase in the mixing matrix for complex $\lambda$ leading to Dirac-type $CP$ violation. The possible measurement of the effective Majorana mass $M_{ee}$ in the neutrinoless double beta decay searches will provide information on the Majorana-type $CP$ violating phase. The analysis of $M_{ee}$ is significant because a stringent constraint on the value of effective Majorana neutrino mass $M_{ee}$ was obtained in the $^{76}Ge$ Heidelberg-Moscow experiment \cite{13} $|M_{ee}| < 0.35 $eV. There are a large number of forthcoming projects such as SuperNEMO \cite{14}, CUORE \cite{15}, COURICINO \cite{16} and GERDA \cite{17} which aim to achieve a sensitivity below 0.01eV. The effective Majorana mass which determines the rate of neutrinoless double beta decay is given by

\begin{figure}
\begin{center}
\epsfig{file=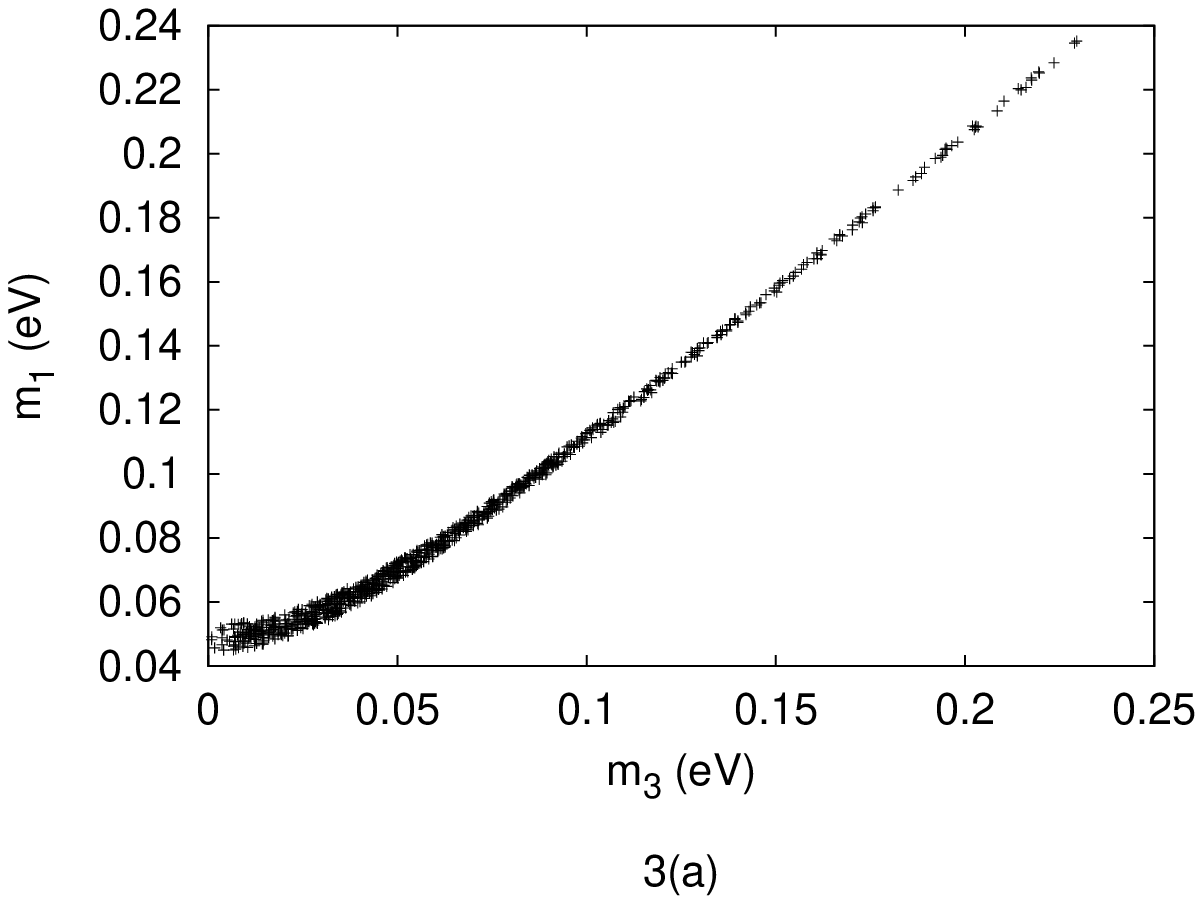,height=5.0cm,width=5.0cm}
\epsfig{file=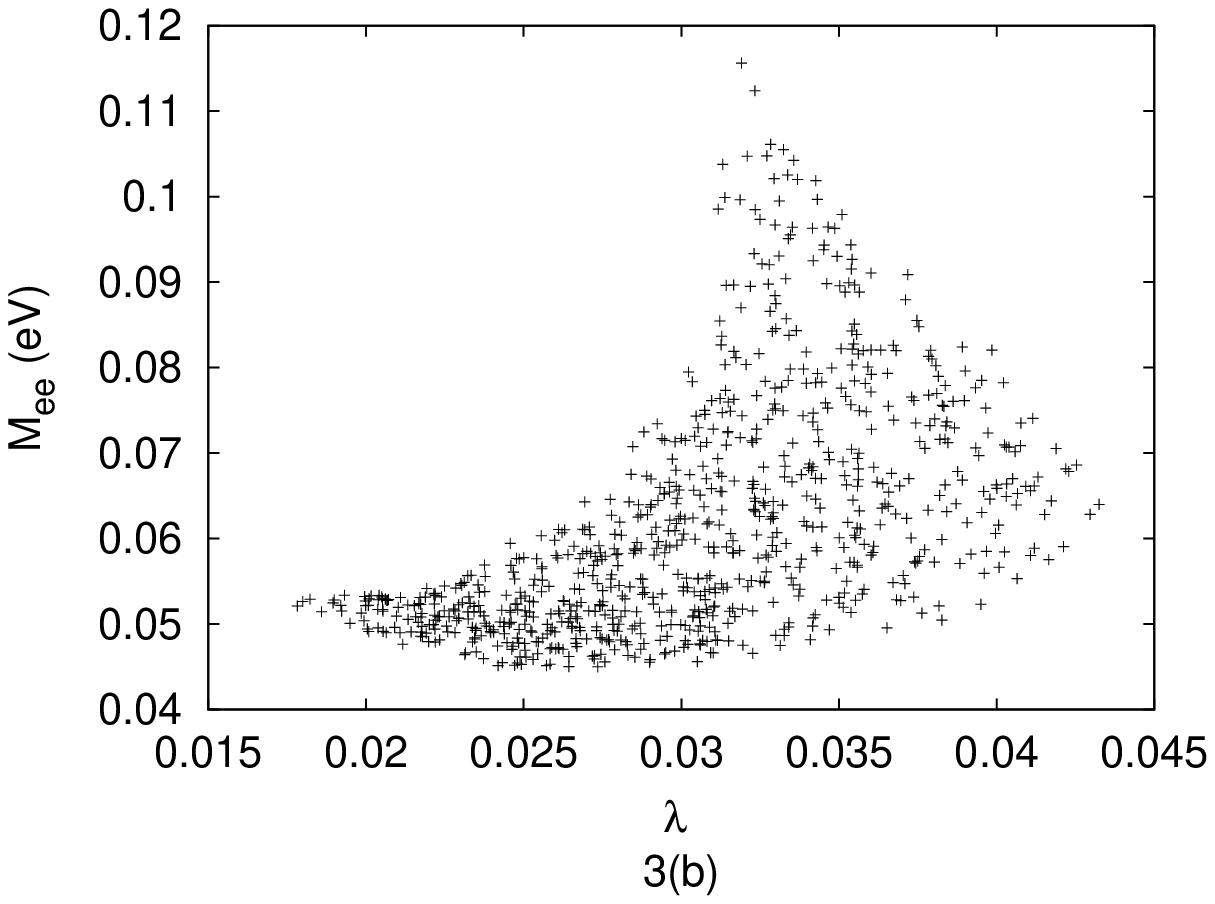,height=5.0cm,width=5.0cm}
\end{center}
\caption{Correlation plots for the inverted hierarchy.}
\end{figure}
\begin{equation}
M_{ee}=|m_1U_{11}'^2+m_2U_{12}'^2+m_3U_{13}'^2  |.
\end{equation}
The value of $M_{ee}$ for normal and inverted hierarchies are $(0.015-0.083)eV$ and $(0.04-0.12)eV$. The correlation plots between the effective Majorana mass and the perturbation parameter $\lambda$ is given in Fig. 2(b) and 3(b).
\section{Summary}
We studied the breaking of $S_3$ invariant neutrino mass matrix in the flavor basis by perturbing it to the third order of a small perturbation parameter to examine its consequences for neutrino masses and mixings. The neutrino mass matrix at the zeroth order gives two degenerate eigenvalues $m_1$ and $m_3$ thus leading to results contrary to the experimental data. This problem is fixed by the introduction of the Majorana phase which renders the magnitude of all the three eigenvalues equal at the zeroth order. This degeneracy is broken by the introduction of perturbation matrices chosen to be $S_3$ group matrices invariant under three different $S_2$ subgroups of $S_3$. The higher order perturbations are suppressed by increasing powers of the small perturbation parameter. The quantum mechanical degenerate perturbation theory uniquely chooses the first order perturbation matrix while the second and the third order perturbations are simultaneously added. The mass-squared differences are found in terms of the perturbation parameter $\lambda$, the free parameter $x$ and parameter $\mu$ which is of the order one. Interestingly, the mixing matrix and the mixing angles in the model are independent of the parameters of the $S_3$ invariant neutrino mass matrix and are determined only in terms of perturbation parameter $\lambda$. This scheme of $S_3$ breaking results in very small deviations from the TBM when confronted with experimental inputs of the mass-squared differences. Both normal and inverted hierarchical mass spectra are allowed in this model. The cause of $CP$ violation in this scenario is the Majorana-type $CP$ violating phase. The allowed range of the effective Majorana mass obtained in the present study lies  well within the reach of the forthcoming experiments providing a stringent test of this model.

\textbf{\textit{\Large{Acknowledgements}}}

The research work of S. D. is supported by the University Grants
Commission, Government of India \textit{vide} Grant No. 34-32/2008
(SR). R. R. G. acknowledge the financial support provided by the Council for Scientific and Industrial Research (CSIR), Government of India.


\begin{thebibliography}{99}
\bibitem{1} P. F. Harrison, D. H. Perkins and W. G. Scott, \textit{Phys. Lett.} \textbf{B 530}, 167 (2002), hep-ph/0202074; Zhi-zhong Xing, \textit{Phys. Lett.} \textbf{B 533}, 85 (2002), hep-ph/0204049; P. F. Harrison and W. G. Scott, \textit{Phys. Lett.} \textbf{B 535}, 163 (2002), hep-ph/0203209.
\bibitem{2} E Ma and G Rajasekaran, \textit{Phys. Rev.} \textbf{D 64}, 113012 (2001); G. Altarelli and F. Feruglio, \textit{Nucl. Phys.} \textbf{B 720}, 64 (2005); \textit{Nucl. Phys.} \textbf{B 741}, 215 (2006); C. S. Lam, \textit{Phys. Rev. Lett.} \textbf{101}, 121602 (2008); for a recent review and list of references, see e.g. G. Altarelli and F. Feruglio, \textit{Rev. Mod. Phys.} \textbf{82}, 2701 (2010), hep-ph/1002.0211
\bibitem{3} L. Wolfenstein, \textit{Phys. Rev.} \textbf{D 18}, 958 (1978); S. Pakvasa and H. Sugawara, \textit{Phys. Lett.} \textbf{B 73}, 61 (1978); \textbf{82}, 105 (1979); E. Durman and H. S. Tsao, \textit{Phys. Rev.} \textbf{D 20}, 1207 (1979); M. Fukugita, M. Tanimoto and T. Yanagida, \textit{Phys. Rev.} \textbf{D 57}, 4429 (1998); H. Fritzsch and Z. Z. Xing, \textit{Phys. Rev.} \textbf{D 61}, 073016 (2000); E. Ma, \textit{Phys. Rev.} \textbf{D 61}, 033012 (2000); M. Tanimoto, \textit{Phys. Lett.} \textbf{B 483}, 417 (2000); E. Ma and G. Rajasekaran, \textit{Phys. Rev.} \textbf{D 64}, 113012 (2001); P. F. Harrison and W. G. Scott, \textit{Phys. Lett.} \textbf{B 557}, 76 (2003); S. L. Chen, M. Frigerio and E. Ma, \textit{Phys. Rev.} \textbf{D 70}, 073008 (2004); F. Caravaglios and S. Morisi, arXiv: hep-ph/0503234; W. Grimus and L. Lavoura, \textit{J. High Energy Phys.} \textbf{08}, 013 (2005); R. N. Mohapatra, S. Nasri and H. B. Yu, \textit{Phys. Lett.} \textbf{B 639}, 318 (2006); N. Haba, K. Yoshioka, \textit{Nucl. Phys.} \textbf{B 739}, 254 (2006); M. Picariello, \textit{Int. J. Mod. Phys.} \textbf{A 23}, 4435 (2008); Y. Koide, \textit{Eur. Phys. J.} \textbf{C 50}, 809 (2007); A. Mondragon, M. Mondragon and E. Peinado, \textit{Phys. Rev.} \textbf{D 76}, 076003 (2007;); C. Y. Chen, L. Wolfenstein, \textit{Phys. Rev.} \textbf{D 77}, 093009 (2008); F. Feruglio and Y. Lin, \textit{Nucl. Phys.} \textbf{B 800}, 77 (2008); M. Mitra, S. Choubey, \textit{Phys. Rev.} \textbf{D 78}, 115014 (2008), hep-ph/0806.3254; Zhi-zhong Xing, Deshan Yang, Shun Zhou, \textit{Phys. Lett.} \textbf{B 690}, 304 (2010); E. Derman, \textit{Phys. Rev.} \textbf{D 19}, 317 (1979); H. Ishimori, T. Kobayashi, H. Ohki, H. Okada, Y. Shimizu and M. Tanimoto, \textit{Prog. Theor. Phys. Suppl.} \textbf{183}, 1-163, (2010) and references therein.
\bibitem{4} C. S. Lam, \textit{Phys. Rev.} \textbf{D 78}, 073015 (2008); Duane A. Dicus, Shao-Feng Ge, Wayne W Repko , \textit{Phys. Rev.} \textbf{D 82}, 033005 (2010).
\bibitem{5} R. Zora, S. Nasri and J. Schechter, \textit{Int. J. Mod. Phys.} \textbf{A 21}, 5875 (2006); R. Zora, J. Schechter and M. Naeem Shahid, \textit{Phys. Rev.} \textbf{D 80}, 093007 (2007); Renata Zora, Joseph Schechter and M. Naeem Shahid, \textit{Phys. Rev.} \textbf{D 82}, 053006 (2010) hep-ph/1006.3307
\bibitem{6} H. Fritzsch, Z. Z. Xing \textit{Phys. Lett.} \textbf{B 440}, 313 (1998), hep-ph/9808272; H. Fritzsch, Z. Z. Xing \textit{Phys. Lett.} \textbf{B 598}, 237 (2004), hep-ph/0406206
\bibitem{7} H. Fritzsch, Z. Z. Xing \textit{Phys. Lett.} \textbf{B 372}, 265 (1996), hep-ph/9509389
\bibitem{8} W. Rodejohann, Z. Z. Xing \textit{Phys. Lett.} \textbf{B 601}, 176 (2004), hep-ph/0408195.
\bibitem{9} S. L. Adler, \textit{Phys. Rev.} \textbf{D 59}, 015012 (1999); erratum 099902, hep-ph/9806518. 
\bibitem{10} P. F. Harrison and W. G. Scott, \textit{Phys. Lett.} \textbf{B 557}, 76 (2003).
\bibitem{11} C. S. Lam, \textit{Phys. Lett.} \textbf{B 640}, 260-262 (2006); S. Kumar, \textit{Phys. Rev.} \textbf{D 82}, 013010 (2010), hep-ph/1007.0808
\bibitem{12} G. L. Fogli \textit{et al.}, \textit{Phys. Rev.}
\textbf{D 78}, 033010 (2008), hep-ph/0805.2517
\bibitem{13} H. V. Klapdor- Kleingrothaus, \textit{Nucl. Phys. Proc. Suppl.} \textbf{145}, 219 (2005). 
\bibitem{14} R. Arnold \textit{et al.}, \textit{Eur. Phys. J} \textbf{C 70}, 927-943 (2010).  
\bibitem{15} C. Arnaboldi \textit{et al.}, 2004a \textit{Nucl. Instrum. Methods Phys. Res., Sect.} \textbf{A 518}, 775 (2004).
\bibitem{16} C. Arnaboldi \textit{et al.}, (CUORICINO collaboration), \textit{Phys. Lett. B} \textbf{584}, 260 (2004).
\bibitem{17} I. Abt \textit{et al.}, (GERDA collaboration) hep-ex/ 0404039.



\end{thebibliography}
\end{document}